\documentclass[traditabstract,letter]{aa} 

\usepackage{graphicx}
\usepackage{natbib}
\usepackage{epstopdf}
\usepackage{txfonts}
\begin{document}

 \title{The onset of solar cycle 24}
 \subtitle{What global acoustic modes are telling us} 
 
    \author{D. Salabert\inst{1,2}
          \and
          R.~A. Garc\'ia\inst{3}
          \and
          P.~L. Pall\'e\inst{1,2}
          \and
          S.~J. Jim\'enez-Reyes\inst{1,2}   
                    }

   \institute{Instituto de Astrof\'isica de Canarias,  E-38200 La Laguna, Tenerife, Spain\\
              \email{salabert@iac.es}
         \and
         Departamento de Astrof\'isica, Universidad de La Laguna, E-38205 La Laguna, Tenerife, Spain     
         \and
         Laboratoire AIM, CEA/DSM-CNRS, Universit\'e Paris 7 Diderot, IRFU/SAp, Centre de Saclay, F-91191 Gif-sur-Yvette, France 
           }

   \date{Received xxxx; accepted xxxx}

   \abstract
     {We study the response of the low-degree, solar p-mode frequencies to the unusually extended minimum of solar surface activity since 2007. A total of 4768 days of observations collected by the space-based, Sun-as-a-star helioseismic GOLF instrument are analyzed. A multi-step iterative maximum-likelihood fitting method is applied to subseries of 365 days and 91.25 days to extract the p-mode parameters. Temporal variations of the $l=0$, 1, and 2 p-mode frequencies are then obtained from April 1996 to May 2009. While the p-mode frequency shifts are closely correlated with solar surface activity proxies during the past solar cycles, the frequency shifts of the $l=0$ and $l=2$ modes show an increase from the second half of 2007, when no significant surface activity is observable. On the other hand, the $l=1$ modes follow the general decreasing trend of the solar surface activity. The different behaviours between the $l=0$ and $l=2$ modes and the $l=1$ modes  can be interpreted as different geometrical responses to the spatial distribution of the solar magnetic field beneath the surface of the Sun. The analysis of the low-degree, solar p-mode frequency shifts indicates that the solar activity cycle~24 started late 2007, despite the absence of activity on the solar surface. }

     
   \maketitle
   
%
\section{Introduction}
The success of helioseismology is due to its capability to accurately measure the p-mode parameters of the solar eigenmode spectrum, which allow us to infer unique information about the internal structure and dynamics of the Sun from its surface all the way down to the core. 
It has greatly contributed to a better understanding of the Sun and provided insights into the complex solar magnetism, through for instance the variability of the characteristics of the p-mode spectrum.

Evidence of low-degree p-mode frequency changes with solar activity, first revealed by \citet{wood85}, were established by \citet{palle89} with the analysis of helioseismic observations spanning the complete solar cycle~21 (1977--1988). Later on, \citet{wood91} measured similar variations in intermediate and high angular-degree modes. As longer, higher quality, and continuous helioseismic observations became available, the solar p-mode frequencies proved to be very sensitive to the solar surface activity not only during solar cycle 23 \citep[see for example,][and references therein]{howe02,chano04} but also over the last three solar cycles (21, 22, and 23), spanning a period of thirty years \citep{chaplin07}, with high levels of correlation with solar surface activity proxies.  

The frequency shifts have also been shown to be angular-degree dependent \citep{chano01}, or rather mode-inertia dependent, but \citet{howe02} found that the latitudinal distribution of the frequency differences for high $l$ modes shows actually close temporal and spatial correlation with the surface magnetic field distribution.
A marginal but significant $l$ dependence of the frequency shifts of the low-degree p modes has also been clearly uncovered \citep{chano04,chaplin04}. The size of the shift also reflects differences in mode inertia. However, the change in inertia with $l$ is small for low-degree modes, and the spatial distribution dominates as well.

The unusual long and deep solar activity minimum as well as the lack of a visible onset of the new cycle~24 have motivated the present work in order to investigate the response of the low-degree p-mode frequencies to this extended minimum (2007 till present) as compared with the rest of cycle~23 (since 1996). 
  
\section{Data and analysis}

\subsection{Data sets}
Observations collected by the space-based instrument Global Oscillations at Low Frequency (GOLF) instrument onboard the {\it Solar and Heliospheric Observatory} ({\it SOHO}) spacecraft were used for this work. GOLF is a resonant scattering spectrophotometer measuring the Doppler wavelength shift -- integrated over the solar surface -- in the D$_1$ and D$_2$ Fraunhofer sodium lines at 589.6 and 589.0~nm respectively \citep{gabriel95}. 
A total of 4768-day velocity time series \citep{garcia05} starting on 1996 April 11  were used  \citep[see calibration method in][]{chano03}. Contiguous subseries of 365 and 91.25 days were produced with a four-time overlap. The series with a filling factor less than 60\% for the 365-day series and 70\% for the 91.25-day series were not used in the following analysis, resulting in a total of 49 and 197 non-independent time series of 365 days and 91.25 days respectively. Thus, the averaged duty cycles for the 365-day and 91.25-day time series were 94.9\% and 97.8\% respectively.

We have also considered contemporaneous observations collected by the ground-based Mark-I instrument at the Observatorio del Teide \citep{brookes78}. Mark-I is also a resonant scattering spectrophotometer but observing in the potassium Fraunhofer line at 769.9~nm. A total of 19 independent yearly time series of the best consecutive 108 days (mean duty cycle 38\%) were used.

\subsection{Mode parameter extraction}
The power spectrum of each 365-day and 91.25-day time series was fitted to yield
estimates of the mode parameters. This fitting was performed by a multi-step iterative method \citep{gelly02,salabert07}. The asymmetric profile of \citet{nigam98} was used to describe each component, as:

\begin{equation}
P_{n,l}(x) = H_{n,l} \frac{(1+\alpha_{n,l} x_{n,l})^2+\alpha_{n,l}^2}{1+ x_{n,l}^2} + B,
\label{eq:mlemodel}
\end{equation}

\noindent
where $x_{n,l} = 2(\nu-\nu_{n,l})/\Gamma_{n,l}$, and $\nu_{n,l}$, $\Gamma_{n,l}$, and $H_{n,l}$ represent the  central frequency, the Full-Width-at-Half-Maximum (\textsc{fwhm}), and the power height of the spectral density respectively for a given mode ($n,l$). The peak asymmetry is described by the parameter $\alpha_{n,l}$, while $B$ represents an additive, constant background level in the fitted window.
Due to their close proximity in frequency, modes are fitted in pairs (i.e., $l=2$ with 0, and $l=3$ with 1). While each mode parameter within a pair of modes are free, the peak asymmetry is set to be the same within pairs of modes.
The mode parameters were extracted by maximizing a likelihood function following the $\chi^2$ with 2 degrees-of-freedom statistics of the power spectrum. The natural logarithms of the mode height, \textsc{fwhm}, and background noise were varied which result in normal distributions. The formal uncertainties on each parameter were then derived from the inverse Hessian matrix. A similar analysis was applied to the Mark-I spectra.

\subsection{Determination of the frequency shifts}
The temporal variations of the p-mode frequencies were obtained by comparing each measured ($l,n$) frequency with the corresponding ($l,n$) averaged frequency over the whole set of analyzed spectra for the different datasets. The formal uncertainties returned by the fits were used as weights in the average computation. The frequency shifts thus obtained were then averaged over the frequency range 2000~$\mu$Hz $\leq \nu \leq 3300~\mu$Hz. Note that due to its lower signal-to-noise ratio, we did not use the $l=3$ modes in the following analysis.
Linear regressions between the frequency shifts and solar surface activity proxies were performed. Mean values of daily measurements of the 10.7-cm radio flux and the Mount Wilson Magnetic Plage Strength Index (MPSI) were calculated over the same 365-day and 91.25-day subseries. 

\begin{figure*}[t]
 \centering
 \includegraphics[angle=90,width=0.8\textwidth]{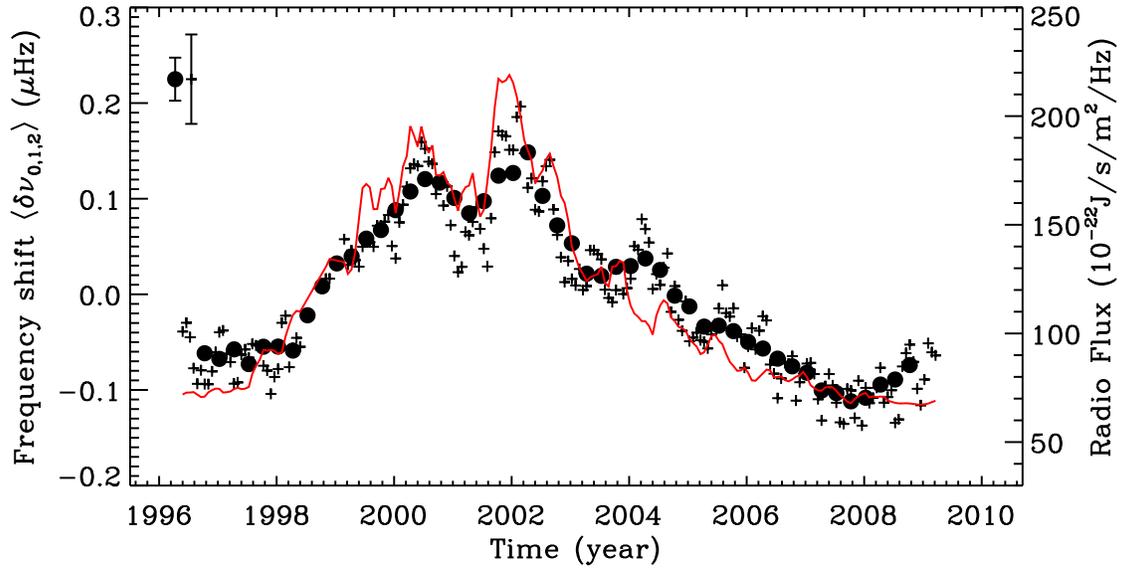} 
  \caption{Mean frequency shifts $\langle \delta\nu_{0,1,2} \rangle$ in $\mu$Hz of the low-degree solar p modes extracted from the analysis of the 365-day (filled circles) and the 91.25-day (plus signs) GOLF spectra.  The associated mean error bars are also represented. The corresponding 10.7-cm radio flux averaged over the same 91.25-day timespan is shown as a proxy of the solar surface activity (solid line).}
 \label{fig:fshift}
 \end{figure*}

\section{Results }
 \begin{figure*}
 \centering
 \includegraphics[width=0.4\textwidth,angle=90]{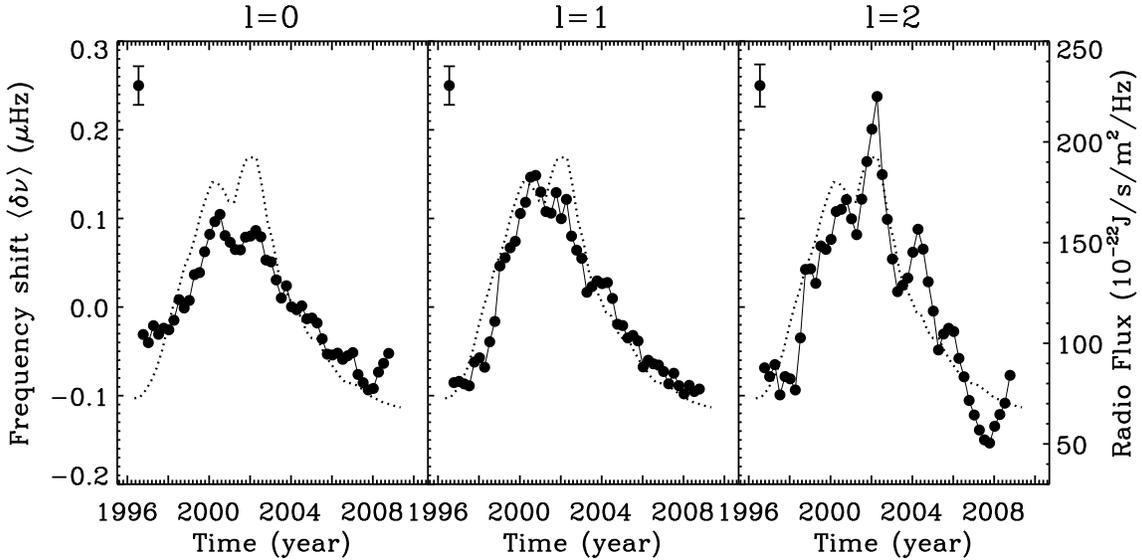}
 \caption{Mean frequency shifts $\langle \delta\nu\rangle$ in $\mu$Hz of the $l=0$,  $1$, and $2$ solar p modes  (left to right panels) extracted from the analysis of the 365-day GOLF spectra. The associated mean error bars are also represented. The corresponding 10.7-cm radio flux averaged over the same 365-day timespan is shown as a proxy of the solar surface activity (dotted line).}
 \label{fig:fshift012}
 \end{figure*}

As the minimum of solar cycle~23 has been longer than predicted and that no surface activity is being observed since the end of 2007, we might expect that the frequency shifts would also show an {\it extended minimum}. Figure~\ref{fig:fshift} shows the temporal variation of the frequency shifts $\langle \delta\nu_{0,1,2} \rangle$ averaged over the modes $l=0$, 1, and 2 extracted from the analysis of both the 365-day and 91.25-day GOLF spectra. As already widely reported, the frequency shifts represented on Fig.~\ref{fig:fshift}  are well correlated with the solar surface activity over cycle~23 (1996--2007). However, as we can see on Fig.~\ref{fig:fshift}, the mean frequency shifts start increasing from the second half of 2007 while no significant surface activity was observed. Excellent qualitative agreement was obtained with the Mark-I data, detailed work being in progress.

Figure~\ref{fig:fshift012} shows the dependence of the frequency shifts $\langle \delta\nu_l \rangle$ on the angular degree $l$ with time, measured from the 365-day spectra. The frequency shifts at each $l$ per unit of change in two global solar activity proxies, as well as the associated Pearson correlation coefficient, $r_p$, the Spearman rank correlation coefficient, $r_s$, and the two-sided significance of its deviation from zero, $P_s$ are given in Table~\ref{table:correl}. These values were obtained from the analysis of the 91.25-day spectra by using only independent points (43 for the period April 1996--June 2007, and 7 for the period June 2007--May 2009). Note that they correspond to one combination of independent series and that they are consistent within $1\sigma$ with any other chosen sets. Given the range of analyzed frequencies and the considered time spans, the amount of the uncovered shifts over cycle~23 and their correlations are consistent within the uncertainties with \citet{chano04} and \citet{chaplin04}. The $l=2$ modes have larger variations in comparison to the $l=1$ and $l=0$ modes respectively.

\begin{figure*}
 \centering
 \includegraphics[angle=90,width=0.9\textwidth]{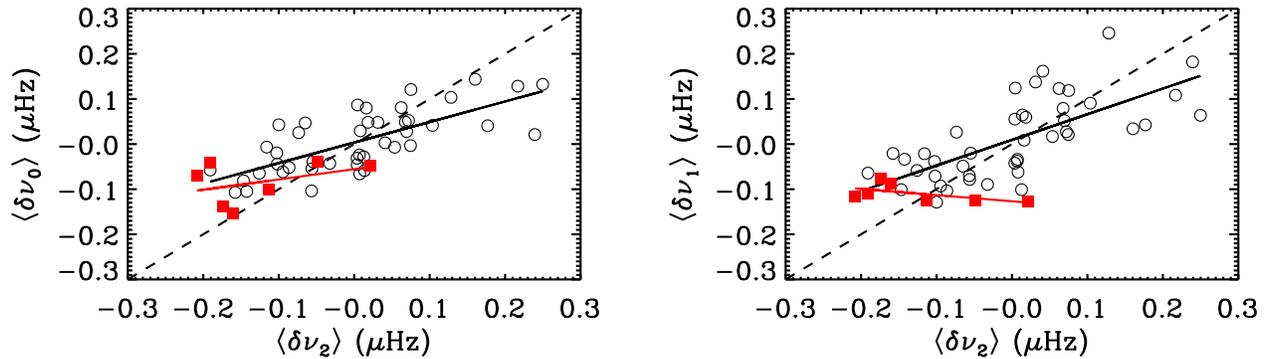} 
  \caption{Relation of the mean frequency shifts (in $\mu$Hz) between angular degree $l$ for the periods April 1996--June 2007 (open circles) and June 2007--May 2009 (filled squares) from the analysis of the 91.25-day GOLF spectra (only independent points were used). Left panel: $\langle \delta\nu_{0} \rangle$ against $\langle \delta\nu_{2} \rangle$. Right panel: $ \langle \delta\nu_{1} \rangle$ against  $\langle \delta\nu_{2} \rangle$. The solid lines correspond to linear fits between the data, and the dashed lines represent the $1:1$ correlation. }
 \label{fig:linear}
 \end{figure*}

From the end of 2007, the frequency shifts of the $l=1$ modes keep slowly decreasing, following the trend of the solar surface activity, while the $l=0$ and $l=2$ modes have a very different behaviour. Indeed, the modes $l=0$ and 2 show increasing frequency shifts, such as expected if solar cycle~24 has already started to progress. The $l=0$ and 2 frequency shifts show negative correlation with surface activity, while the $l=1$ frequency shifts are in some extent still correlated with surface activity (Table~\ref{table:correl}). Note that the significances of having null correlation are larger because of much fewer number of independent points (7 points) for the period June~2007--May~2009. In order to stress this different behaviour with $l$, Fig.~\ref{fig:linear} compares the frequency shifts of the $l=0$ and $l=1$ modes in relation with those of the $l=2$ modes for the two periods April~1996--June~2007 and June~2007--May~2009, confirming what is shown on Fig.~\ref{fig:fshift012}. 
While the modes present similar behaviour during cycle 23 up to the first half of 2007, the $l=0$ and 2 modes behave differently compared to the $l=1$ modes from June~2007 and onwards.

The different behaviours observed between values of  angular degree can be interpreted in terms of the spatial geometry of the modes measured with the GOLF instrument. Indeed, Sun-as-a-star observations such as the ones made by the GOLF and Mark-I instruments are only sensitive to the modes with $l+|m|$ even. This is because the solar rotation axis lies close to the plane of the sky from the observing point. The observed $l=1$ frequency is therefore a weighted measurement of the visible components ($l=1, |m|=1$), while in the case of the $l=2$ mode, the fitted frequency corresponds to the weighted measurement of the zonal ($l=2, m=0$) and sectoral ($l=2, |m|=2$) components \citep{chano04}. The sectoral modes are more concentrated along the equator while the zonal modes are most sensitive to the high latitudes. 
Thus, the $l=2$ modes seen by full-disk instruments are more sensitive to the high latitudes of the Sun. Moreover, they have larger frequency variations than the $l=0$ modes, which average the entire visible solar disk, while the full-disk $l=1$ mode is more concentrated along the equator. The results presented here would then indicate that the process responsible for the frequency shifts at high latitudes and related to the new solar cycle~24 started late 2007.

\section{Summary and discussion}

\begin{table*}
\begin{minipage}{\textwidth}
\caption{Frequency shifts at each $l$ per unit of change in solar proxy and correlations from the analysis of independent 91.25-day GOLF spectra. }
\label{table:correl}      
\centering                 
\renewcommand{\footnoterule}{}  
\begin{tabular}{l r c c r r r r c}      
\hline\hline                
       &  \multicolumn{4}{c}{April 1996--June 2007}   &   \multicolumn{4}{c}{June 2007--May 2009}           \\    
$l$ & \multicolumn{1}{r}{Gradient\footnote{Gradient against the radio flux in units of nHz RF$^{-1}$ (with 1~RF =  10$^{-22}$~J~s$^{-1}$~m$^{-2}$~Hz$^{-1}$); against the MPSI in units of nHz G$^{-1}$.}} & \multicolumn{1}{c}{$r_p$} & \multicolumn{1}{c}{$r_s$} & \multicolumn{1}{c}{$P_s$} & \multicolumn{1}{c}{Gradient$^a$} & \multicolumn{1}{c}{$r_p$} & \multicolumn{1}{c}{$r_s$} & \multicolumn{1}{c}{$P_s$} \\ \hline
       
 &  \multicolumn{8}{c}{ \it{10.7-cm radio flux} }        \\  
  
   0 & 1.3 $\pm$ 0.2 & 0.82 &  0.77 & $1.15 \times 10^{-9}$ & $-16.6$ $\pm$ 12.1     &  $-0.66$ & $-0.71$ & 0.07 \\  
   1 & 1.7 $\pm$ 0.2 & 0.82 &  0.82 & $1.93 \times 10^{-11}$ & 6.2 $\pm$ 11.5         &  0.52       & 0.43       & 0.34 \\
   2 & 2.1 $\pm$ 0.2 & 0.86 &  0.85 & $5.55 \times 10^{-13}$ & $-26.7$ $\pm$ 13.1 & $-0.48$  & $-0.32$ & 0.48 \\
   
    &  \multicolumn{8}{c} {\it{Mount Wilson MPSI} }          \\
    
    0 & 57.4 $\pm$ 7.3 & 0.81 & 0.78 & $9.74 \times 10^{-10}$ & $-498.3$ $\pm$ 440.6 &  $-0.58$ & $-0.39$ & 0.38 \\  
   1  & 78.9 $\pm$ 7.0 & 0.83 & 0.85 & $7.37 \times 10^{-13}$ & 239.6 $\pm$ 419.1       & 0.56        & 0.46       & 0.29 \\
   2  & 96.0 $\pm$ 7.8 & 0.85 & 0.84 & $2.63 \times 10^{-12}$ & $-827.3$ $\pm$ 457.1 & $-0.44$  & $-0.07$ & 0.88 \\

\hline         
\end{tabular}
\end{minipage}
\end{table*}

While solar p-mode frequency shifts have been shown to be closely correlated with the solar surface activity over the past solar cycles \citep{howe02, chano04, chaplin07}, it appears that during the unpredicted long   minimum of cycle~23, the frequency shifts of the low-degree p modes show unexpected behaviour. Since the second half of the year 2007, the frequency shifts are observed to be uncorrelated with surface activity. While no significant surface activity was observed, the frequency shifts of the modes $l=0$ and 2 show significant increases and those of the $l=1$ modes follow the extended minimum. The difference between the $l=0$ and 2 modes and the $l=1$ modes can be interpreted as a different response to the spatial distribution of the solar magnetic field. In Sun-as-a-star measurements such as the GOLF observations, the modes $l=0$ and 2 modes are more sensitive to the high latitudes than the $l=1$ modes. Such a behaviour of the frequency shifts would indicate variations in the magnetic flux at high latitudes related to the onset of solar cycle~24. This result qualitatively agrees with the measurements of the cosmogenic isotopes \element[][10]Be \citep{beer98} and \element[][14]C \citep{stuiver89} which show an 11-year periodic variation even during extremely low epochs of surface activity, such as  the Maunder minimum. 

Also, to estimate in a simple manner when the minimum of cycle 23 occurred, we performed a second order polynomial fit to the mean frequency shifts ($l=0$ and 2) over the period 2006--2009. To do so, we used only independent points from the analysis of the 91.25-day series. Thus, we obtained that the minimum of solar cycle 23 must have happened during the last quarter of 2007.

The prediction of the solar cycle properties, such as its strength and pattern, give significantly different conclusions for cycle~24 \citep[see e.g.,][]{svalgaard05,dikpati06}. Most of these models are calibrated using solar observable, such as the sunspot number, which reflect changes in the solar surface. It is clear that there is a time delay between the emergence of magnetic fields located probably close to the tachocline and the appearance of activity on the surface. The physical conditions beneath the surface are thus modified during this process. Our findings show that the signature of activity appear before in the solar acoustic mode parameters than in other solar activity proxies, and therefore must be taken into account in the calibration of the solar dynamo models.

In order to study in more details the spatial dependence of the frequency shifts during this peculiar solar minimum, this work will be extended to the analysis of the spatially-resolved observations from the Global Oscillation Network Group (GONG) and Michelson Doppler Imager (MDI) instruments, which allow the decomposition into individual $m$ components. \citet{chano04} previously showed that the values of the frequency shifts for $l=0$, 1, and 2 measured in MDI data scale to the corresponding spherical harmonic components of the observed line-of-sight surface magnetic field. The analysis of the variability of the other p-mode parameters during this extended minimum is also currently underway. 
 
\begin{acknowledgements}
The authors thank O.~L. Creevey for helpful discussions and useful comments on the manuscript.
The GOLF instrument onboard SOHO is a cooperative effort of many individuals, to whom we are indebted. SOHO is a project of international collaboration between ESA and NASA. We thank all the members, past and present, of the helioseismology group at the IAC for doing Mark-I observations and maintenance. The use of Birmingham University resonant scattering spectrophotometer at Observatorio del Teide is also deeply acknowledged. The 10.7-cm radio flux data were obtained from the National Geophysical Data Center. This study includes data from the synoptic program at the 150-Foot Solar Tower of the Mt. Wilson Observatory. The Mt. Wilson 150-Foot Solar Tower is operated by UCLA, with funding from NASA, ONR and NSF, under agreement with the Mt. Wilson Institute. D. S. acknowledges the support of the grant PNAyA2007-62650 from the Spanish National Research Plan. This work was supported by the European Helio- and Asteroseismology Network (HELAS), a major international collaboration funded by the European Commission's FP6, and by the CNES/GOLF grant at the SAp/CEA-Saclay.
\end{acknowledgements}



\end{document}